\documentstyle[spie]{article} 
\input{psfig}   
\newcommand{\sas}{{\it XMMSAS}}
\newcommand{\xmm}{{\it XMM-Newton}}
\newcommand{\rgs}{{\it RGS}}
\newcommand{\mos}{{\it MOS}}
\newcommand{\pn}{{\it pn}}
\newcommand{\epic}{{\it EPIC}}
\newcommand{\abel}{Abell~209}
\newcommand{\etal}{et~al.} 
\newcommand{\fig}[1]{Fig.~\ref{#1}}
\newcommand{\tab}[1]{Tab.~\ref{#1}}
\newcommand{\cha}[1]{\S\ref{#1}}
\newcommand{\eqn}[1]{Eqn.~\ref{#1}}

\title{Data analysis methods for \xmm\ observations of extended sources.\\
       Application to bright massive clusters of galaxies at $z=0.2$} 

\author{Philippe B. Marty\supit{a,b},\\
        Jean-Paul Kneib\supit{b}, Rachida Sadat\supit{b},
        Harald Ebeling\supit{c}, Ian Smail\supit{d}
\skiplinehalf 
\supit{a}I.A.S., b\^at. 121, Campus Universit\'e Paris 11, F-91405 Orsay cedex, France 
\\
\supit{b}L.A.O.M.P., 14 av. E. Belin, F-31400 Toulouse, France
\\
\supit{c}Institute for Astronomy, University of Hawaii, 2680 Woodlawn Drive, Honolulu, Hl.96822, USA
\\
\supit{d}Department of Physics, University of Durham, South Road, Durham DH13LE, UK
}

\authorinfo{Further author information: (Send correspondence to P.B.M.)\\P.B.M.: E-mail: marty@ias.fr}

\begin{document}
\maketitle

\begin{abstract}
	In this paper, a review is given of methods useful for \xmm\ \epic\ data analysis of extended sources, along with some applications to a bright massive cluster of galaxies at $z=0.2$, \abel.
	This may constitute an introduction to that kind of advanced analysis, complementing cookbooks that can be found over the internet, the \xmm\ \epic\ calibration status document\cite{bib:Kirsch02}, and the data analysis workshops organised in VILSPA, which were only dedicated so far to point sources analysis.
	In addition, new spectro-imaging techniques are proposed, in order to measure for instance the intracluster medium mass and temperature profiles, or even maps.
\end{abstract}



\section{INTRODUCTION}
\label{sec:intro} 

	After two years and a half of in-orbit operations, the \xmm\ telescopes have almost completed their GT (Guaranted Time) and GO1 (1st Guest Observer cycle) duties. The selection process among the proposals for the GO2 (2nd Guest Observer cycle) is about to end, so that the corresponding observations may start within a semester, by early 2003. \tab{tab:clgstat} shows that ClG (Clusters of Galaxies) related proposals represents one third to one seventh of the total, depending on whether one includes connected topics (individual galaxies, groups of galaxies and large scale structures surveys) or not. It may be assumed that the same ratio will appear in the final GO2 schedule, and has already appeared in the GT and GO1 schedules. However, \tab{tab:clgstat} indicates a rather lower ratio in terms of published results.

\begin{table*}[h!bt]
\caption{\label{tab:clgstat}Scheduling and bibliographical overview of Clusters of Galaxies as seen by the \xmm\ \epic\ instruments.}
\begin{center}
\begin{tabular}{|p{0.2\textwidth}|p{0.7\textwidth}|}
\hline
PV/CAL$^a$+GT+GO1 &
	at least 129 scheduled$^b$ ClG targets \\
&	at least 77 observed$^b$ ClG at the time of REV 421 (03/2002) \\
\hline
GO2$^c$ &
	113 proposals about ClG \\
&	115 proposals about galaxies and groups \\
&	 43 proposals about surveys \\
&	869 overall proposals \\
\hline
after 2.5 years$^d$ &
	18 papers$^e$ about ClG: 16 objects \\
&	+7 papers about M87 \\
&	+2 papers about serendipitous detections \\
&	+2 papers about preliminar results from surveys \\
&	at least 250 to 300 papers about \xmm\ observations in general \\
\hline
\end{tabular}
\end{center}
\footnotesize
  \begin{list}{}{}
    \item[$^{\mathrm{a}}$] Phase of Verification and CALibration
    \item[$^{\mathrm{b}}$] Source: \verb!http://xmm.vilspa.esa.es/!
    \item[$^{\mathrm{c}}$] Source: \xmm\ internet news \#22
    \item[$^{\mathrm{d}}$] Source: internet ADS (Abstracts Database Service); including {\it astro-ph} preprints when not elsewhere published yet
    \item[$^{\mathrm{e}}$] including 3 in common with {\it Chandra}
  \end{list}
\normalsize
\end{table*}


	As a matter of fact, clusters of galaxies, along with {\it supernovae} remnants, are the only X-rays diffuse emitters that may cover a substantial fraction of the, if not the whole, \xmm\ telescope FOV (Field Of View). The case of the diffuse EXRB (Extragalactic X-Rays Background) is more complex since it is proven\cite{bib:Mus00} to originate from the cumulated emission of an unresolved distribution of distant AGN (Active Galaxy Nuclei), in correlation\cite{bib:Pug96} with the FIR (Far Infra-Red) background. But at any rate, this kind of observations suffers from the main \xmm\ advantage, which can turn into a drawback, namely its great throughput and sensitivity. With its spectro-imaging capability extending upto $12~keV$, as compared to that of the previous major X-rays observatory {\it ROSAT}, limited to $3~keV$, and its three mirrors combined effective area $3$ to $5$ times greater than that of the contemporary {\it Chandra} telescope\cite{bib:XMMUHB}, \xmm\ opens the window of spatially resolved spectroscopy, and achieve photon statistics just sufficient to detect fainter parts of extended objects of which only the cores were previously analysed. But the problems araising then are those of any pioneer: common data analysis tools evolved so as to encompass these new capabilities, but are obviously limited in dealing with all the new parameters and corrections necessary; uncertainties in instrument calibration may reach the same order of magnitude than photon statistics for faint regions falling on the edges of the FOV (where the mirror vignetting effect is important), so that it is not always clear whether extending the analysis upto those faintest parts of an extended source is limited either by the accuracy of the models fitted to the data, or by the calibration corrections applied to the data (or both).

	In the following sections (\cha{sec:bkg},\cha{sec:spectro}), a review is presented of the main analysis methods used so far in the frame of \xmm\ \epic\ observational data, with a list of their advantages and drawbacks, as well as proposals for improving them, and examples applied to the analysis of a bright massive ClG, namely \abel. But before going into the data analysis details, the next section (\cha{sec:detec}) summarizes briefly the main instrumental effects that will have to be dealt with further on.

\section{INSTRUMENTAL ASPECTS}
\label{sec:detec}

  \subsection{Photon detection efficiency}

	An \xmm\ X-rays telescope basically consists of a chain of 3 optical devices: (a) the mirror module which collects and focuses the light; (b) a filter (chosen among the six self-explanatory different positions of a filter wheel: ``open'', ``thin1'', ``thin2'', ``medium'', ``thick'', ``closed'') aimed at blocking UV and greater wavelength and reducing the X-rays flux from the brightest sources; (c) a CCD camera which detects the photons through the photo-electric effect, and is able to function in a photon counting mode due to the high amount of energy deposited by each single photon. Two of the three telescopes are also equipped with a reflection grating device which allows for dispersing half of the flux collected by the mirrors onto a high resolution spectrometer (\rgs\ instrument) at a secondary focal plane.

	Each of these devices has its own reflection or transmission or detection efficiency, the combination of which forms the overall telescope efficiency. A basic description of their origin and consequences is given hereafter, more details and graphs may be found in the \xmm\ user's handbook\cite{bib:XMMUHB}, other dedicated papers\cite{bib:Asch00,bib:Struder01,bib:Turner01} and other articles {\it ibid}.

    \subsubsection{Mirrors effective area and vignetting}

	The basic reflectivity of the grazing angles mirrors is known to decrease with energy, as a consequence of the absorbtivity of its constitutive materials (here Gold and Nickel) and of the Bragg's law which states the optimal incidence angle as a function of the wavelength: $2 ~ d \cdot cos(\theta_i) = n \cdot \lambda$.

	This coefficient, between $1$ and $0$, may be multiplied by the nominal collecting surface (in $cm^2$) of the mirror module to form the effective area.

	Another effect results from the fact that some X-rays may fall in the mirrors FOV with a direction not parallel to the optical axis, for instance those coming from secondary point sources on the edge of the FOV, or those coming from the edge of very extended objects while the telescope axis is aligned on its core. These off-axis X-rays hence have angles of incidence slightly off the optimal Bragg value and suffer a loss of reflectivity; the more off-axis, the less efficient. Again, this effect, known as vignetting, is increasing (the reflectivity is decreasing) with energy.

	We may thus summarize the mirror overall efficiency as follow:
\begin{equation}
\label{eqn:mirror}
	\mbox{ME}(E,\theta) = A_{eff}(E) \cdot V(E,\theta)
\end{equation}

    \subsubsection{Filter transmission}

	The transmission here is simply a function of the thickness of the absorbing layers (mainly Aluminium and Polypropylen or Polyamide). The efficiency hence increases with the photon energy and decreases with the thickness of the filter. One may expect small spatial inhomogeneities according to a possible spatial variation of the thickness, but this is assumed to be negligeable and anyway has not been accurately measured yet.

	We have then:
\begin{equation}
\label{eqn:filter}
	\mbox{FE}(E) = T(E)
\end{equation}

    \subsubsection{Detector quantum efficiency}

	When the remaining photons, after going through the mirror and the filter, eventually hit the camera, there are many ways of losing some of the photo-electrons created by the impact: missing hits (because the photon were not enough energetic to penetrate the semiconductor, or too energetic and went completely through without depositing all its energy), scattering within neighbouring pixels (``patterns''), mixing with neighbouring photons (``pile up''), loss during photo-charges transfer to the electronic readout node (``charge transfer inefficiency'' or CTI)\ldots

	This results in two main effects. On one hand, the overall number of counted photo-electric events within a given lapse of time is lower than the incident photon flux, due to an overall efficiency lower than $1$; this may be expressed in terms of a detection QE (Quantum Efficiency) as a function of the incident photon energy. On the other hand, there is a slight redistribution effect of the energy spectrum of the incident photon flux, due to various charge transfer and electronic readout mechanisms which may lead to some photo-events being affected an energy value lower than that of the incident photon; this may be represented by a two-dimensional matrix\cite{bib:Arnaud98}, the RMF (Redistribution Matrix Function), which is function of the incident photon energy and the output electronic channel energy.

	We then define the detection QE and RMF as follow:
\begin{eqnarray}
\label{eqn:detect}
\mbox{QE}(E) & = & N_{counts}^{detector} \over N_{photons}^{detector}(E) \\
\label{eqn:redist}
\mbox{RMF}(E,C) & = & N_{counts}^{detector}(C) \over \mbox{QE}(E) \cdot N_{photons}^{detector}(E)
\end{eqnarray}

    \subsubsection{Overall response functions}

	Assuming the detector is operated in pile-up limiting conditions (increasing the filter thickness and decreasing the camera reading cycle in order to observe sources with increasing flux), one can define\cite{bib:Arnaud98} an ``on-axis ARF'' (Ancilliary Response Function) by grouping \eqn{eqn:mirror}, \eqn{eqn:filter} and \eqn{eqn:detect} with $\theta = 0$:
\begin{eqnarray}
\label{eqn:ARF}
\mbox{ARF}(E) & = & \mbox{ME}(E,0) \cdot \mbox{FE}(E) \cdot \mbox{QE}(E) \\
              & = & N_{counts}^{detector} \over N_{photons}^{sky,\theta=0}(E)
\end{eqnarray}

	In the same way, one can combine the RMF and the ARF (\eqn{eqn:redist} and \eqn{eqn:ARF}) to form the overall telescope on-axis response matrix:
\begin{eqnarray}
\label{eqn:RMF}
\mbox{Resp}(E,C) & = & \mbox{RMF}(E,C) \times \mbox{ARF}(E) \\
             & = & N_{counts}^{detector}(C) \over N_{photons}^{sky,\theta=0}(E)
\end{eqnarray}

	These response functions may be used (see \cha{sec:spectro}) by some dedicated softwares to fold a given physical X-rays emission spectrum model (along a true energy axis) and try fitting the observed data (which are projected onto a digital units, or channels, energy axis).

  \subsection{Instrument spatial resolution}

	X-rays mirrors, as those of any other telescopes, suffer from a spatial redistribution of the events on the detector, with respect to the incident photon flux, in an analogous fashion as the energy dependent RMF. It is called the PSF (Point Spread Function), and determines the spatial resolution of the instrument. In general the PSF has a 2D-gaussian shape, and the spatial resolution may be defined as the gaussian FWHM (Full Width at Half Maximum); in the case of the \xmm\ mirrors, the PSF is better modelled as King profiles\cite{bib:Ghiz02}, and its FWHM (which would be equal to twice the King profile core radius if the power index were unity) varies between $4$ and $8~arcsec$ across the FOV and the energy passband.

  \subsection{Background origins}

	During quiescent periods (no significant soft protons contamination), the \epic\ background has been shown\cite{bib:Lumb01} to be mainly composed of:
\begin{itemize}
\item Remaining cosmic rays induced events (high energy; non vignetted): unrejected by on-board electronics.
\item Instrumental fluorescent X-rays emission (from camera close environment, mainly the Aluminium line around $1.4~keV$ for both \mos\ and \pn, and the Copper complex around $8~keV$ for \pn; non vignetted): no significant increase in fluorescence level after solar flares has been observed so far, but images in corresponding sharp energy bands show that these emission features are correlated to regions of corresponding material in the camera structure\cite{bib:XMMUHB}.
\item Electronic noise (bright pixels and dark current; non vignetted): (a) since the \pn\ pixels are about the size of the PSF, bright pixels are sometimes undistinguishable from point sources (hence missed by the \sas\ algorithms); (b) an electronic overshoot problem seems to cause noise accumulation near the \pn\ readout edges (outer edge of the FOV); (c) ionizing particles secondary effects on the \mos\ often materialize as flickering pixels at low energies; (d) finally, the \mos\ dark current may be negligeable (of the order of $0.5~cnt/s/CCD$ in the $0.2-10~keV$ band\cite{bib:Ghiz00}) and the same may be expected from the \pn.
\item Remaining low flux soft protons (high energy; vignetted): a magnetic divertor in the telescopes prevents electrons from reaching the cameras, but some protons may still contaminate the data; while this effect is still under investigations, it seems that these protons mainly show up as additional flickering pixels occuring at dates very close to a forthcoming flaring period\cite{bib:Gendre01}.
\item Sky X-rays background (low to medium energy; vignetted): in the case of ClG observations, looking away from the galactic center, these events are mainly due to the EXRB, mostly unresolved faint and/or far AGN sources\cite{bib:Mus00}.
\end{itemize}

\section{BACKGROUND SUBTRACTION} 
\label{sec:bkg}

	The first step before performing any advanced spectro-imaging analysis on X-rays data is to clean them from any pertubating and/or non-X-rays events.

  \subsection{Selecting only valid events}
\label{sec:valev}

	Despite the first electronic processing onboard, a lot of non-photon events are still present in the raw data, and can only be sorted out after a standard pipeline processing (\sas\ tasks: \verb!emchain! or \verb!emproc! for the \mos, \verb!epchain! or \verb!epproc! for the \pn) aiming at pattern and energy recognition, which includes a flagging step where events falling on or near badpixels or CCD edges are marked accordingly. Minimizing onboard processing, despite its necessity for limiting the telemetry, allows for software and algorithms improvements, and old data sets may be re-analysed later with better results and minimal information loss.

	\tab{tab:flags} summarizes the different flag values. By hexadecimal combinations, it is possible to give a multiple description for a given event, if need be. Inversely, by excluding some hexadecimal combinations, one can select a cleaned subset of the original events. A straightforward method consists of keeping only those events having a total flag value equal to $0$ (no warning), by adding to the \sas\ task \verb!evselect! the following filter: \verb!FLAG==0!.

	A more subtle method is proposed by the current pipeline software, which consists of removing the defined flags from number 16 to 31, respectively: \verb!(FLAG & 0x766b0000)==0! for the \mos, \verb!(FLAG & 0xfa0000)==0! for the \pn. Note that, in the latter case, events from out the FOV are still in, which may not be the desired results, so that one should prefer for the \pn: \verb!(FLAG & 0x2fb0000)==0!. 

	Finally, it would seem safe to exclude also events from offset columns and spoiled frames, to remove electronic noise, ending with the filter formulae: \verb!(FLAG & 0x766b0808)==0! for the \mos, \verb!(FLAG & 0x2fb0808)==0! for the \pn.

\begin{table*}[h!bt]
\caption{\label{tab:flags}Events quality FLAG values description.}
\begin{center}
\small
\begin{tabular}{|l|l|l|}
\hline
short code&command line to select    &description \\
\hline
XMMEA\_0  & (FLAG \& 0x        1)!=0 & DIAGONAL \\
XMMEA\_1  & (FLAG \& 0x        2)!=0 & CLOSE TO CCD BORDER \\
XMMEA\_2  & (FLAG \& 0x        4)!=0 & CLOSE TO CCD WINDOW \\
XMMEA\_3  & (FLAG \& 0x        8)!=0 & ON OFFSET COLUMN \\
XMMEA\_4  & (FLAG \& 0x       10)!=0 & NEXT TO OFFSET COLUMN \\
XMMEA\_5  & (FLAG \& 0x       20)!=0 & CLOSE TO ONBOARD BADPIX \\
XMMEA\_6  & (FLAG \& 0x       40)!=0 & CLOSE TO BRIGHTPIX \\
XMMEA\_8  & (FLAG \& 0x      100)!=0 & CLOSE TO DEADPIX \\
XMMEA\_9  & (FLAG \& 0x      200)!=0 & CLOSE TO BADCOL \\
XMMEA\_10 & (FLAG \& 0x      400)!=0 & CLOSE TO BADROW \\
XMMEA\_11 & (FLAG \& 0x      800)!=0 & IN SPOILED FRAME \\
XMMEA\_16 & (FLAG \& 0x    10000)!=0 & OUT OF FOV \\
XMMEA\_17 & (FLAG \& 0x    20000)!=0 & IN BAD FRAME \\
XMMEA\_19 & (FLAG \& 0x    80000)!=0 & COSMIC RAY \\
XMMEA\_20 & (FLAG \& 0x   100000)!=0 & MIP ASSOCIATED (pn) \\
XMMEA\_21 & (FLAG \& 0x   200000)!=0 & ON BADPIX \\
XMMEA\_22 & (FLAG \& 0x   400000)!=0 & SECONDARY (pn) REJECTED BY GATTI (mos)\\
XMMEA\_23 & (FLAG \& 0x   800000)!=0 & TRAILING (pn) \\
XMMEA\_25 & (FLAG \& 0x  2000000)!=0 & OUT OF CCD WINDOW \\
XMMEA\_26 & (FLAG \& 0x  4000000)!=0 & OUTSIDE THRESHOLDS (mos) \\
XMMEA\_28 & (FLAG \& 0x 10000000)!=0 & ON BADROW (mos) \\
XMMEA\_29 & (FLAG \& 0x 20000000)!=0 & BAD E3E4 (mos) \\
XMMEA\_30 & (FLAG \& 0x 40000000)!=0 & UNDERSHOOT (mos) \\
\hline
\end{tabular}
\normalsize
\end{center}
\end{table*}

	In the same way, events originating from piled up photons and/or cosmic rays show different pattern photon events. A description of the patterns may be found in the bibliography\cite{bib:Turner01}, along with the following additionnal selection criterion: \verb!PATTERN in [0:12]! for the \mos\ (singles, doubles, triples and quadruples), \verb!PATTERN in [0:4]! for the \pn\ (singles and doubles only, because the \pn\ pixels are larger). Note that the \mos\ pattern number 31, which could also point at real photo-events, has been excluded since it appears extremely noisy\cite{bib:Ghiz00}.

	Leaving events with non valid flag or pattern value would lead to meaningless results, especially for spectral analysis, where some non valid \pn\ events may have negative energy values, or some special \mos\ events between $12$ and $15~keV$ show a triangular distribution along time since they are triggered by an electronic readout device (GATTI).

	Finally, a third selection criterion on the event energy\footnote{in PI (or Pulse Independent) units; {\it i.e.} electronic channel units translated into equivalent $eV$ units; but remember that, due to redistibution effect, this energy may not be that of the incident photon and still is a kind of ``channel'' energy} may be set up to filter out regions were the spectral response of the \epic\ cameras is not well calibrated and/or rather noisy, respectively: \verb!PI in [300:12000]! for the \mos, \verb!PI in [300:15000]! for the \pn. Nevertheless, on the one hand, the low energy calibrations are improving regularly, and on the other hand, the high energy part is dominated by other sources of noise (\cha{sec:flare}), so that this criterion should become less important in the future. In addition, a last energy band selection may be performed independently with most of the spectral analysis softwares.

  \subsection{Keeping only quiescent periods}
\label{sec:flare}

\begin{figure}[h!bt]
\begin{center}
\hspace{0cm}
\psfig{figure=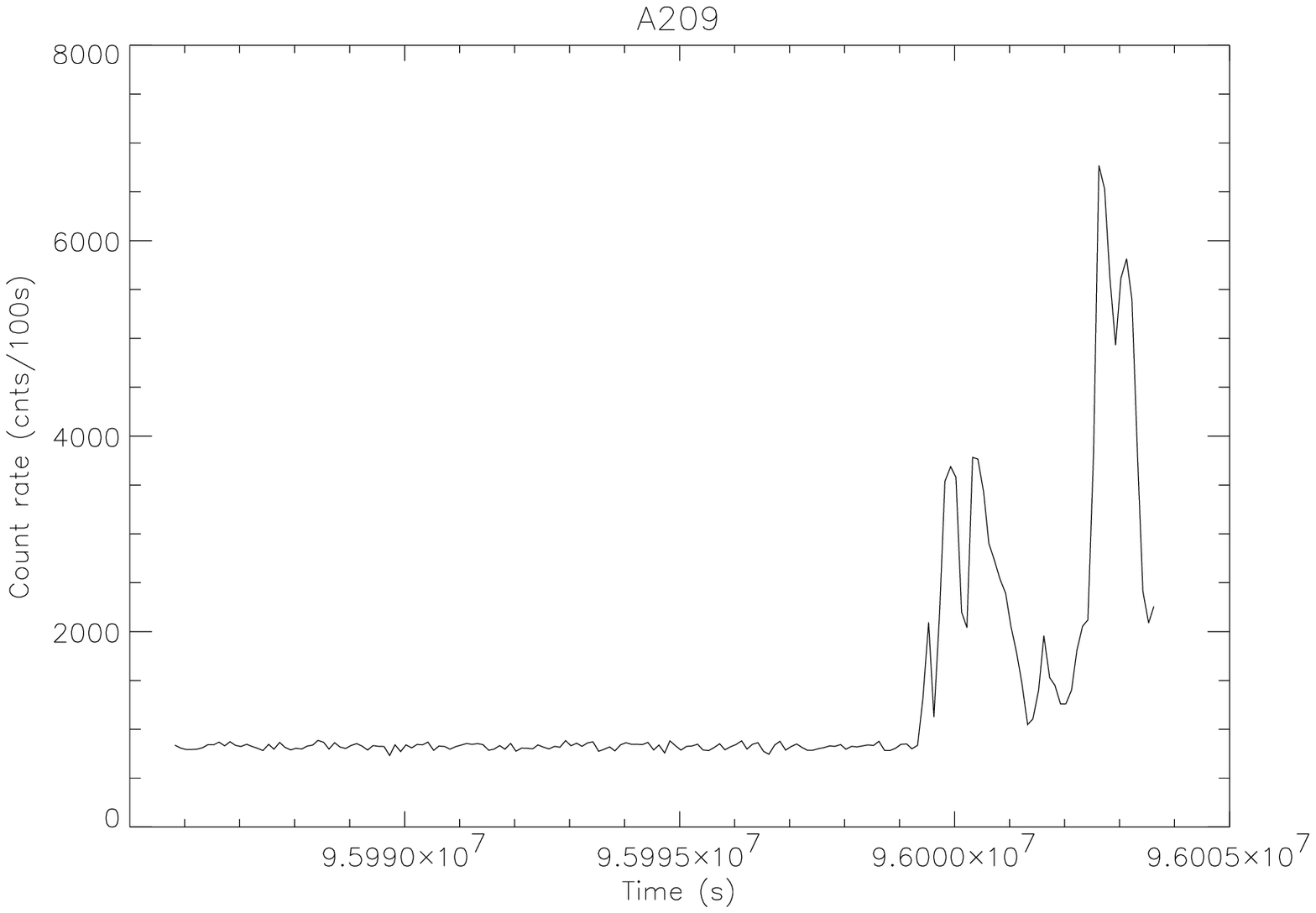,width=0.45\textwidth}
\psfig{figure=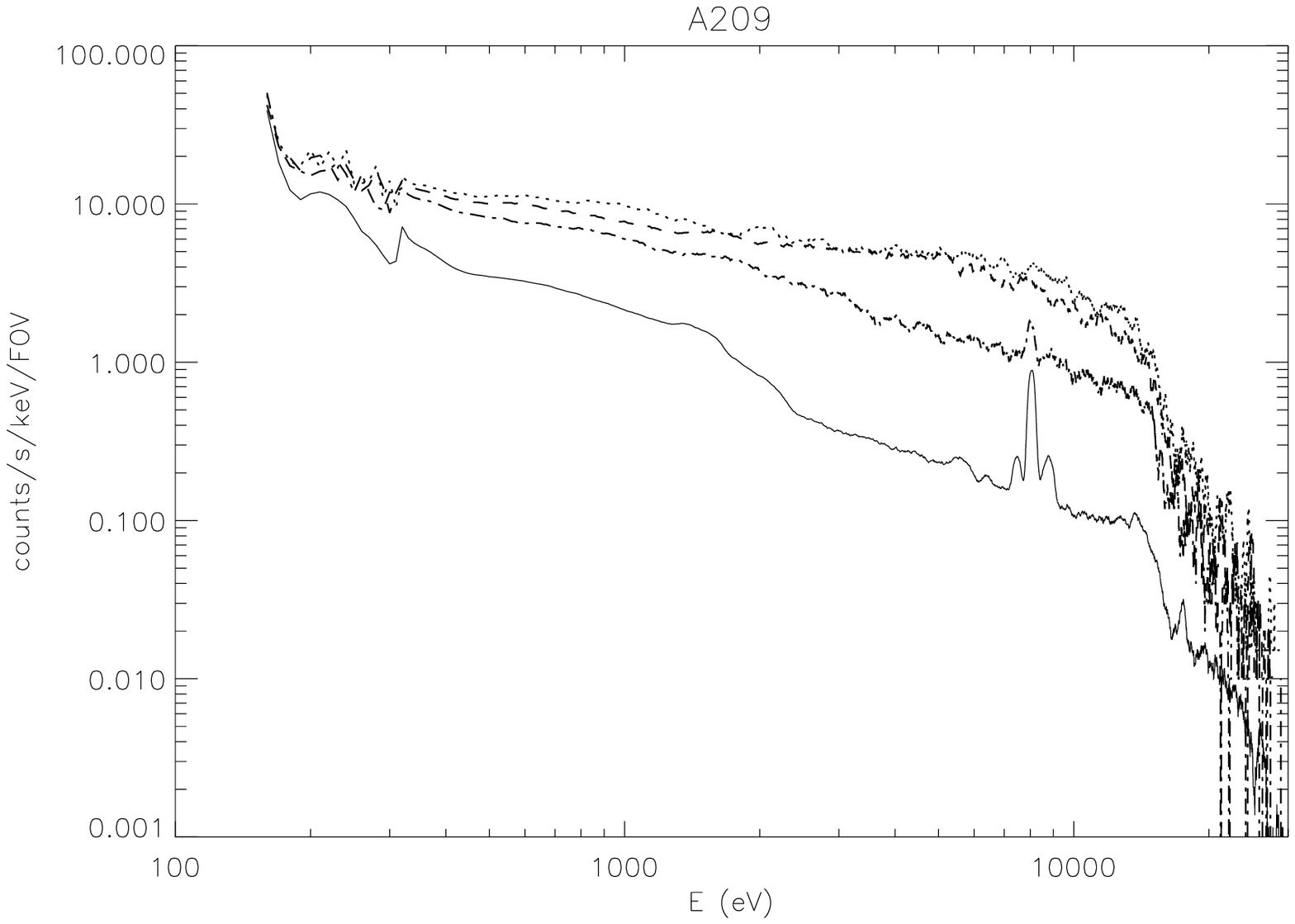,width=0.45\textwidth}
\caption{\label{fig:flare}LEFT: lightcurve of the \pn\ valid events with no energy nor time selection, during the \abel\ observation (REV 202). RIGHT: spectrum of three flaring periods (dash: 96000500-96000800, dash-dot: 96002600-96002800, dot: 96003000-96003300) as compared to that of a quiescent period (95985777-95999000).}
\end{center}
\end{figure}

	Some \xmm\ \epic\ observations presents periods (from $0$ to $100\%$ of the total observation duration\ldots\ generally about $5\%$) of unusually high background level (\fig{fig:flare}, left), which mainly consists of photon-like events, at rather high energies (from $1$ to $20~keV$). These events are thought to originate from solar soft protons being directly gathered by the grazing mirrors modules, and/or maybe trapped beforehand by earth magnetosphere. Studies of correlation with altitude and attitude of the telescope with respect to the sun and the earth magnetic field lines are still under work\cite{bib:Pietsch02}. Their time variation is highly chaotic and their spectral shape also seems to be variable (\fig{fig:flare}, right).

	At least five different methods may be used to screen out these flaring periods, with more or less accuracy but also complexity.

	The most straightforward method is based on a selection of time intervals where the count rate in a given energy band is lower to a given threshold. This is the most robust and the simplest to implement into the \sas: tasks \verb!evselect! with criterion from \cha{sec:flare} to generate a lightcurve, then \verb!tabgtigen! to generate a range of GTI (Good Time Intervals) to be added to further selection operations. However, the energy range and threshold have to be found out empirically and manually. But, in general, values of $10~counts$ per $100~s$ bin for each \mos, or $80~counts$ per $100~s$ bin for the \pn\ work fine in the $200$ to $12000~eV$ range. Other threshold and energy bands may be found in the litterature\cite{bib:Arnaud01,bib:Lumb01,bib:Lumb02,bib:Maj02,bib:Marty01,bib:Marty02,bib:Pratt01}. They mainly depends on whether very hard objects (like neutron stars, which may drastically raise the mean count rate even above $12~keV$) are present in the FOV, and on the local overall background for the considered observation (a few rare data sets show an anormalously high ``quiet background'' level in addition to shorter soft protons flaring periods). A last variant consists of doing exactly the same analysis but only on events situated on detector parts outside the FOV: since the soft protons are mechanicaly diffused within the telescope, they may reach these regions, as well as photons from instrumental fluorescence, while photons from the sky should not. But since these regions are rather limited in area, the statistics is quite poor and this variant should be discouraged.

\begin{figure}[h!bt]
\begin{center}
\hspace{0cm}
\psfig{figure=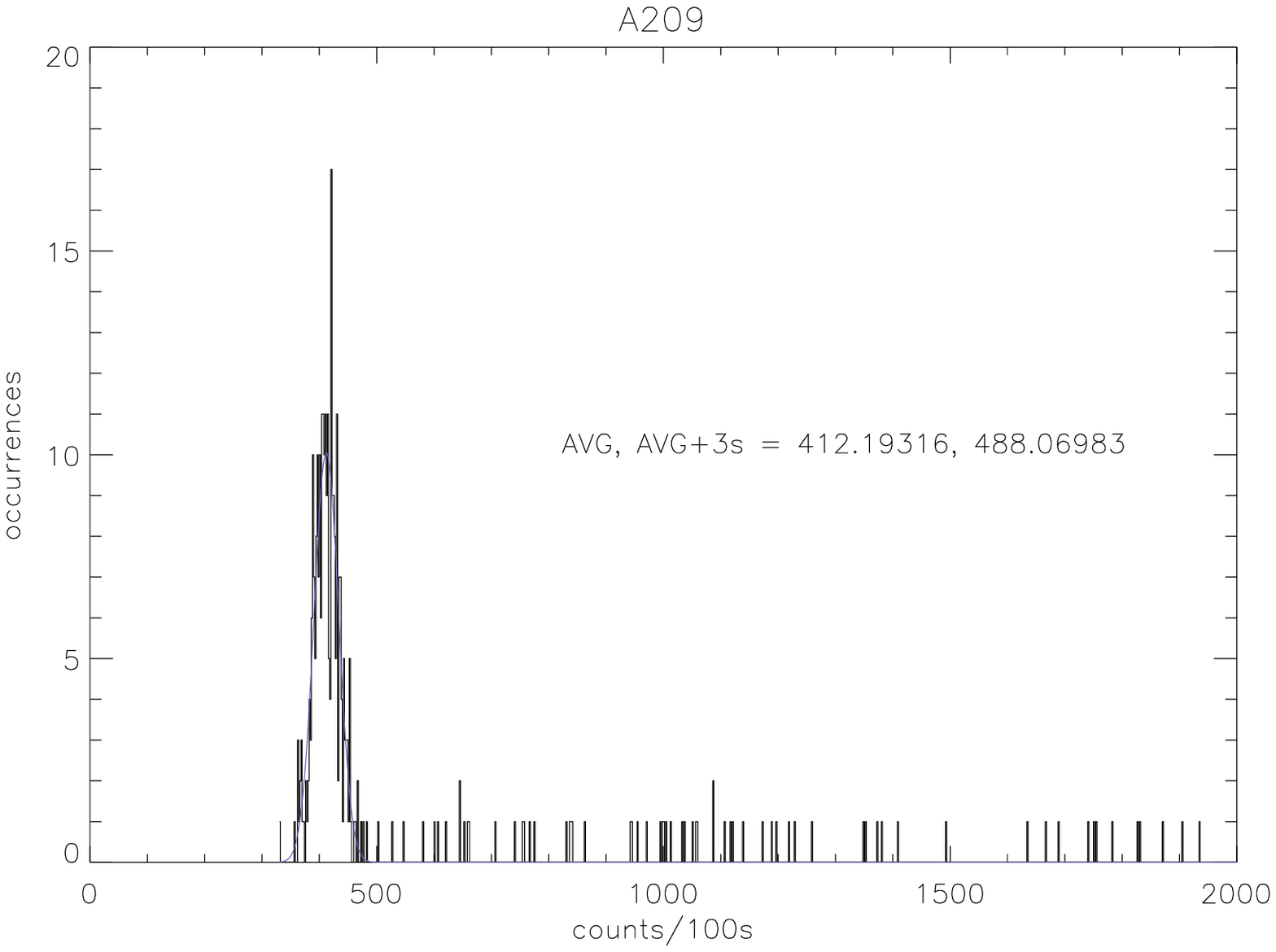,width=0.45\textwidth}
\psfig{figure=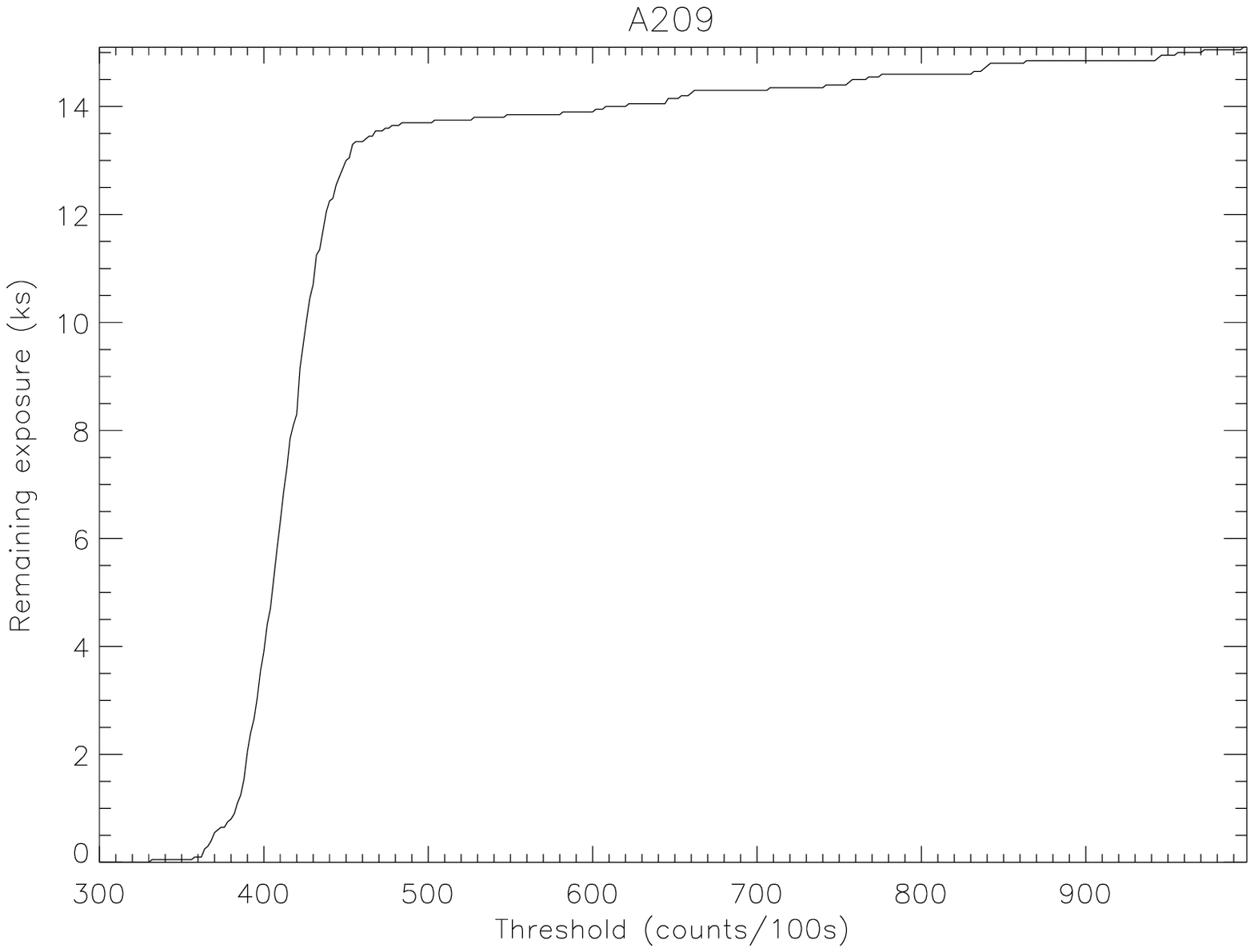,width=0.45\textwidth}
\caption{\label{fig:flare2}LEFT: histogram of the \pn\ valid events lightcurve with no energy nor time selection, during the \abel\ observation (REV 202). RIGHT: plot of the remaining exposure time as a function of the threshold applied to the \pn\ valid events lightcurve.}
\end{center}
\end{figure}

	A second method\cite{bib:Lumb02} is a bit more sophisticated and propose a fixed recipe to determine a count rate threshold, given a starting energy band: to compute the average count rate $\bar{c}$ and choose as a threshold value $\bar{c} + 3\sigma$ (\fig{fig:flare2}, left). This presents a drawback and an advantage: statistical functions provided within the \sas\ (tasks \verb!statgets! or \verb!lcplot!) are not sufficient because the flares peaks will contaminate the moments values, and a gaussian fit over the histogram of the rate curve would be the correct algorithm, but requires use of an external routine, not always easy to implement in a batch analysis; however, in addition to its automatic design, this method has been tested succesfully on data sets contaminated by soft protons upto $75\%$.

	A third method\cite{bib:Maj02} consists of an empirical determination of the threshold value, for a given energy band, by analysing the remaining exposure time after thresholding as a function of the threshold value. This should lead to an asymptotic graph (\fig{fig:flare2}, right), which should allow to determine the optimum threshold. This iterative method appears very slow, even in an automatic batch process, and the asymptotic threshold determination not trivial.

\begin{figure}[h!bt]
\begin{center}
\hspace{0cm}
\psfig{figure=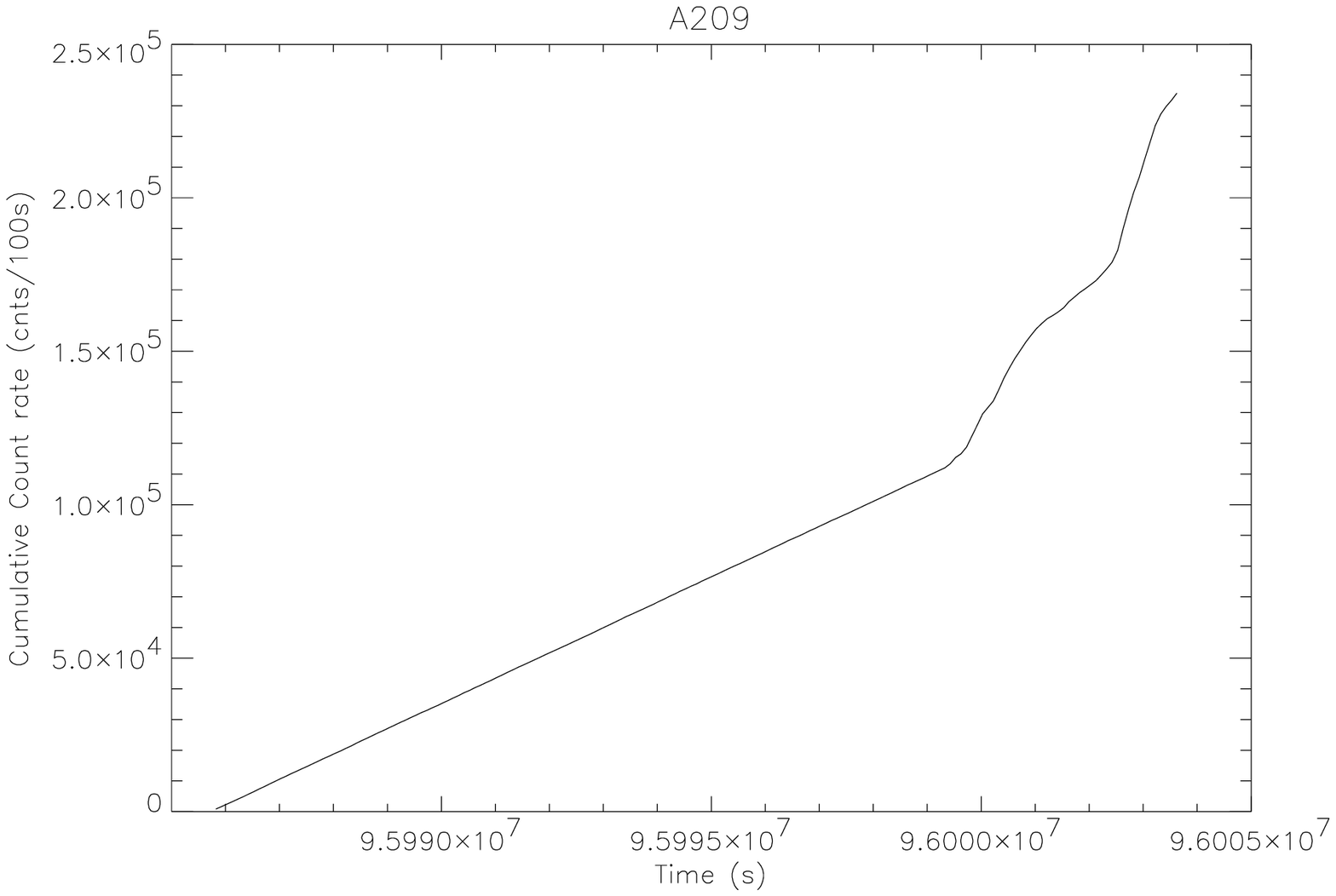,width=0.45\textwidth}
\psfig{figure=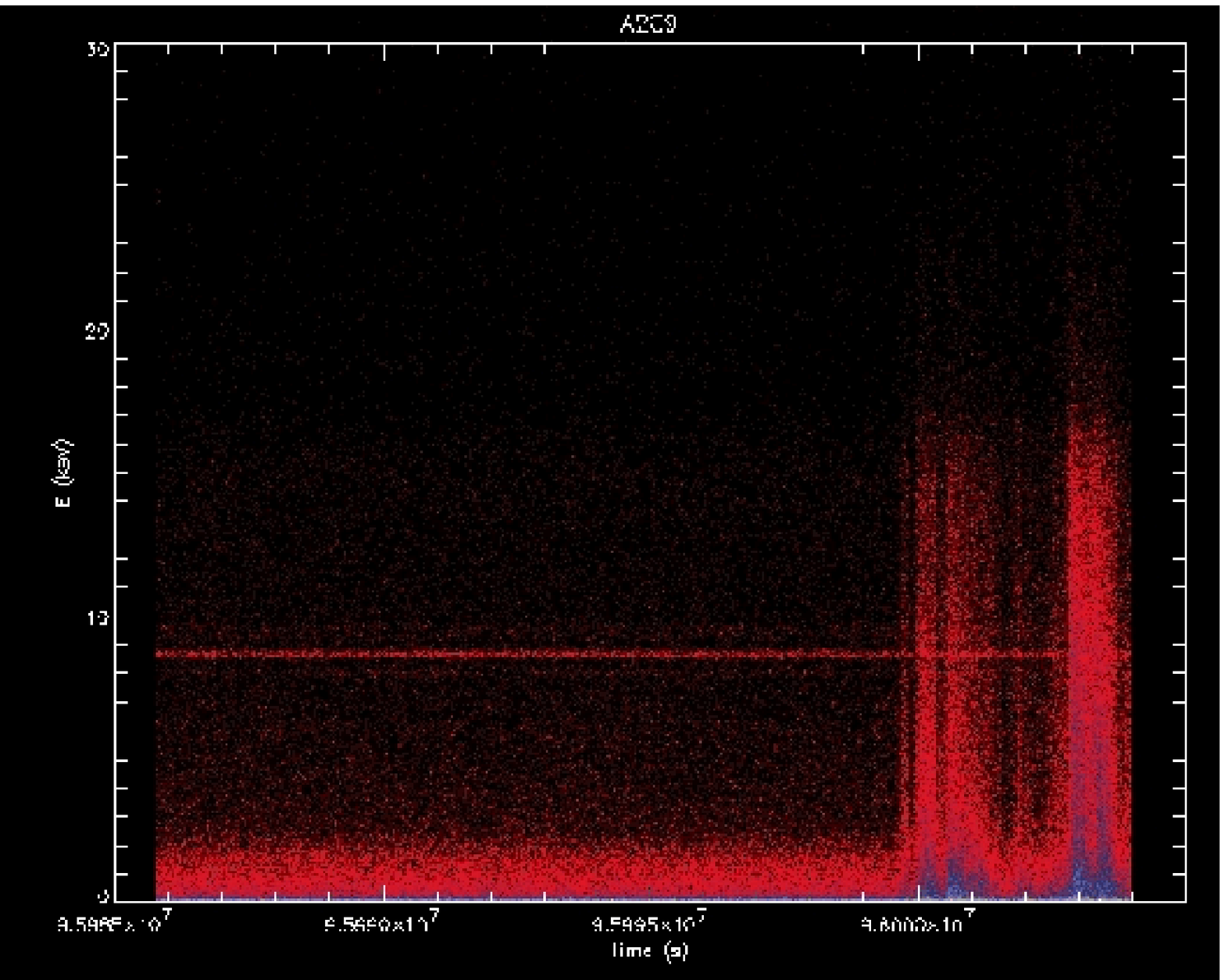,width=0.45\textwidth}
\caption{\label{fig:flare3}LEFT: cumulative lightcurve of the \pn\ valid events with no energy nor time selection, during the \abel\ observation (REV 202). RIGHT: energy-time diagram for the \pn\ valid events.}
\end{center}
\end{figure}

	A fourth method would be based on an analysis of a cumulative lightcurve (a cumulative histogram of the events dates), where quiescent regions would appear as straight lines of equal slope, separated by steeper steps at the dates of proton flaring (\fig{fig:flare3}, left). The detection of linear regions seems easier but still beyond the capabilities of the \sas, hence uneasy to implement in a batch analysis.

	A fifth method would be based on a clustering analysis (relying on wavelet transformations) on an energy-time (\fig{fig:flare3}, right) or space-time diagram, where flares clearly appears as stronger vertical patches. This has not been fully developped yet, but could provide a solution for complete removal of soft protons contamination.

	Remaining events associated to soft protons flares would mainly result in an artificial hardening of the source spectrum, hence leading to higher temperature determination in the case of the intracluster medium. Up to now, remaining flickering pixels due to low level soft protons contamination of quiescent periods cannot be directly detected and removed (see \cha{sec:skybkg}).

  \subsection{Estimating all other background components}
\label{sec:skybkg}

	To remove all other background components (mainly: instrumental fluorescence, remaining cosmic rays and the EXRB), it has been proposed\cite{bib:Arnaud01} to build pseudo-observation data sets from real observation data of ``blank'' sky regions, like the Lockman Hole, and to use these auxiliary data to estimate and remove the background from the main analysed object.

	More details can be found in a dedicated technical note\cite{bib:Lumb01}, but the basic principle is to merge several different data sets in order to achieve the best possible statistics, lowering background error bars below the level of the main data errors. This merged blank set may then be reprojected onto the sky attitude of the main object, so that exactly the same cleaning selection may be performed, hence resulting in the same effective detector area, making any subtraction possible, without any bias. But many other problems araise then.

	First of all, these merged data sets reach important file size, so that they are a trade off between statistics and memory usage. Their size may be reduced by applying cleaning selection (like those in \cha{sec:valev}, \cha{sec:flare}), but this could prevent users to define different criteria and introduce a bias with respect to their main data analysis. In the end, only two selections were applied, the first one being a flare removal as conservative as possible, allowing the user for a second tighter pass if need be.

	Then, any ``blank'' field never is purely ``blank'' and always holds field stars and sources. The second selection was indeed a manual removal of point sources, along with an automatic badpixel screening. With at least five different data sets being merged, it is assumed that ``holes'' thus created in each data set will be filled by the others, so no further correction is needed. But the point is that each data set amounts for at least $30~ks$ exposure, while for most of ClG observations the typical remaining exposure after cleaning rarely reaches $20~ks$. This means that faint sources that are easily resolved in the empty sky data may still lie in the noise in the ClG data, thus leading to an underestimation of the empty sky and hence to an overestimation of the remaining ClG signal, specially at low energies ($< 2~keV$) according to the EXRB and field stars cumulated spectra.

	In addition, these auxiliary blank data sets have only been generated, yet, using observations with the ``thin'' filter on, assuming that it was the most used filter setting during observations of extended sources, which flux are by definition extended enough to limit the risk of pile-up. Unfortunately, many such observations must be carried out with the medium filter due to the closing presence of a bright star in the FOV, and many {\it supernovae} remnants are both extended and very bright sources and also require use of thicker filter. This may have no influence on the pure instrumental noise (fluorescence), and very limited influence on the very energetic noise (cosmic rays) since all filters transmissions tend to unity above $3~keV$\cite{bib:Turner01}, so that it could be expected that subtracting blank data may always remove these components, whatever the filter settings. But a major bias is introduced at low energies concerning the EXRB estimation. If the ClG is observed with thicker filter than the blank fields, this would tend to go in the opposite direction as the previous effect, overestimating the background level relatively to the main data.

	Another bias cause for the low energies resides in the different levels of galactic absorption (due to the hydrogen column density on the line of sight, as measured by the $N_H$) between blank fields, which are chosen as ``blank'' because they have low $N_H$ values as well as limited field source population, whereas typical ClG observations may suffer from $N_H$ values $2$ to $5$ times higher, even worse for some galactic {\it supernovae} remnants objects. This again tends to an overestimation of the EXRB from the blank fields relatively to the main source, and may be aggravated by a filter setting bias.

	Yet another bias related to instrumental settings may appear for the \pn\ camera only. It comes from the fact that the blank fields have mainly been observed in the ``full frame'' mode, while the ``extended full frame'' mode should be prefered in the case of relatively low flux sources, because this mode has a longer readout cycle and hence is more sensitive to pile up, but less to out-of-time events contamination. The \pn\ camera has no memory zone where to stack photo-charges of a previous frame while integrating the next one, nor any obturator to prevent further photons to impact the camera during the readout process. This results in an accumulation of low energy events near the readout nodes, as well as some bright columns full of trailing events in the regions exposed to higher fluxes. The consequence in terms of background estimation within blank fields that have been observed in a different mode is that a spectral discrepancy will appear below $500~eV$.

	Finally, because of various factors, including electronic gain influence, the mean instrumental and energetic cosmic rays background levels inside the blank fields are expected to be slightly different from that within the main data. Many variants of renormalization have been tested (inside or outside the FOV, in different energy bands or selection criteria), but probably the most robust one relies on the comparison of count rates outside the FOV in the whole analysis energy range. Indeed, these regions should only contain instrumental and cosmic rays noise (\cha{sec:flare}), allowing to directly compare background levels between blank fields and main data, while the broader energy range ensures minimal statistics. However, this renormalization finishes to make any EXRB component estimation completely meaningless because there is no reason for it following the same variations than the instrumental or high energies (from solar and/or galactic center origin) components.

	The following sections propose ways to improve this situation. As a last note, it should be noticed that any time- and/or pointing dependent variability in these various background components still needs to be carefully investigated, even if it only occurs at the moment through those huge flarings periods (to be removed), $N_H$ differences (still to be quantified) or small instrumental renormalizations (typical factors values between $0.75$ and $1.3$). Also, the high energy component may not be as insensitive to the filter thickness as it may be expected, since cosmic rays are not photons but energetic nuclei, and transmission through the filter follows different matter interaction laws; for instance, some particles may deposit upto $50\%$ of their energy in the ``thick'' filter, so that the result in terms of photo-charges noise on the camera may be well different than if the particle had gone through the ``thin'' filter. In the same fashion, changing the Aluminium thickness in front of the detector may change the fluorescence level of the Aluminium line without affecting much of the continuum background level, but again this still has to be quantified. On the other hand, blank fields may be expected to hold a similar fraction of remaining low level soft protons so that they may be the only way, yet, to remove this component from the main data.

  \subsection{Isolating the instrumental background}
\label{sec:detbkg}

	It may seem natural now to look for a method allowing to isolate the instrumental and high energy particles background components from the EXRB component. One way has been called the ``double background subtraction'' method\cite{bib:FerDor02}. It relies on the assumption that a detector region may be defined free of any sky sources within the main data. A first direct subtraction of the blank field from the main data, assuming that there is no filter bias nor operating mode bias and that all selections and renormalizations have been made, allows for instrumental and particles noise removal. While the signal from the region holding the analysed source is still pertubated by the EXRB and the biased subtraction (due to possible $N_H$ difference, abusive renormalization and/or unresolved point sources), the signal within the outer ``free'' region should be exactly equal to that perturbation. Hence, after taking into account the difference in terms of effective detector surface, the signal from the outer free region may simply be subtracted again from the main region to get at last a cleaned measurement.

	Another possibility is offered by data sets acquired with the filter wheel in the ``closed'' position. This ensures that no photons from the sky may reach the camera, so that the data set only contains instrumental and particles (which may still go through or induce secondary fluorescence as before) components, like a kind of ``dark field''. A second advantage of these data is that they hold much less events for a given exposure duration than an equivalent blank field, precisely because of the absence of sky photons. As for blank fields, a set of merged dark fields may thus be used to clean the main data, leaving again the EXRB correction for a second pass: either from an outer free region of the main data, or from a model, or even from blank fields which have been also cleaned for their dark components in the same way. This second step is discussed in \cha{sec:spectro}.

\begin{figure}[h!bt]
\begin{center}
\hspace{0cm}
\psfig{figure=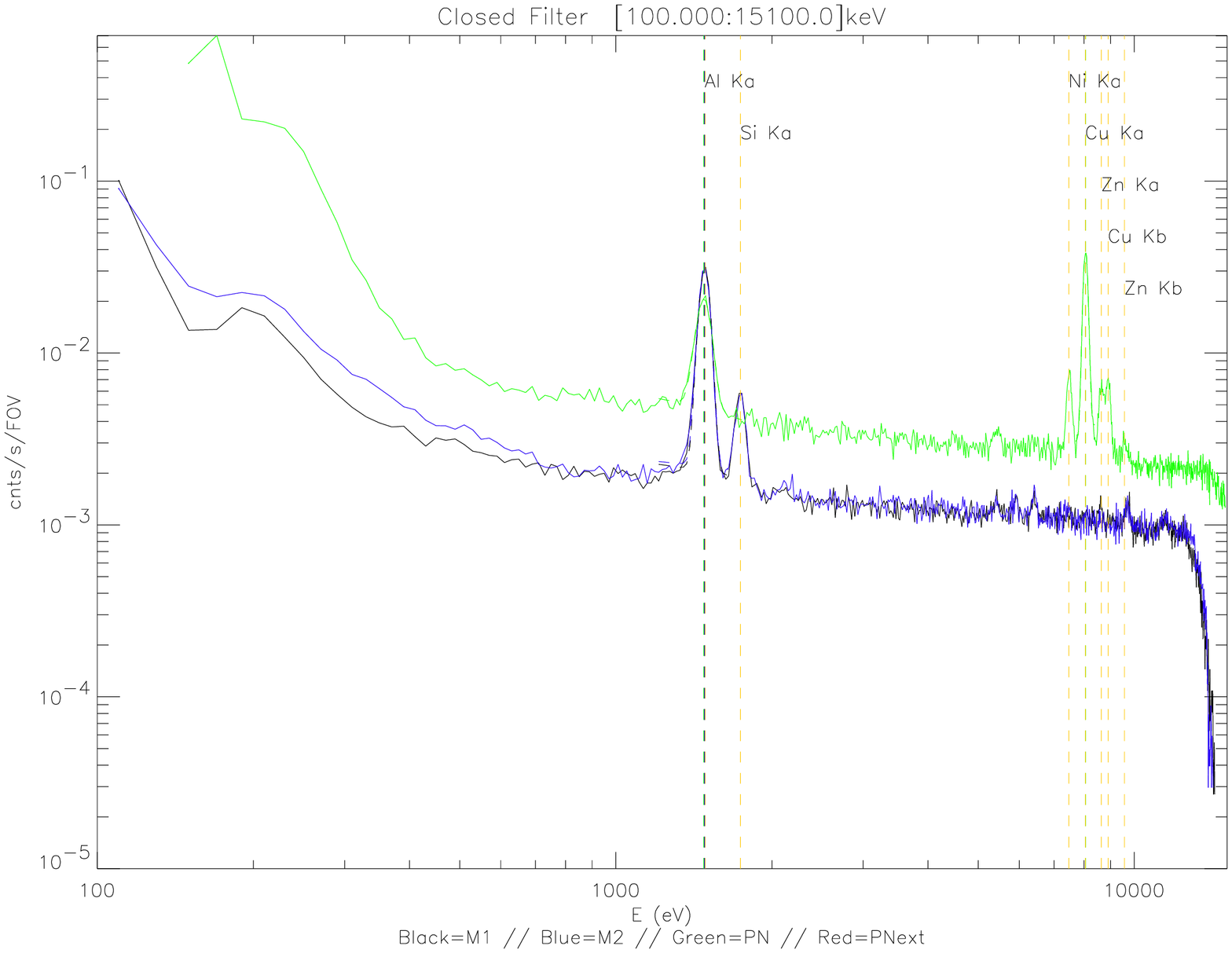,width=0.45\textwidth,angle=90}
\psfig{figure=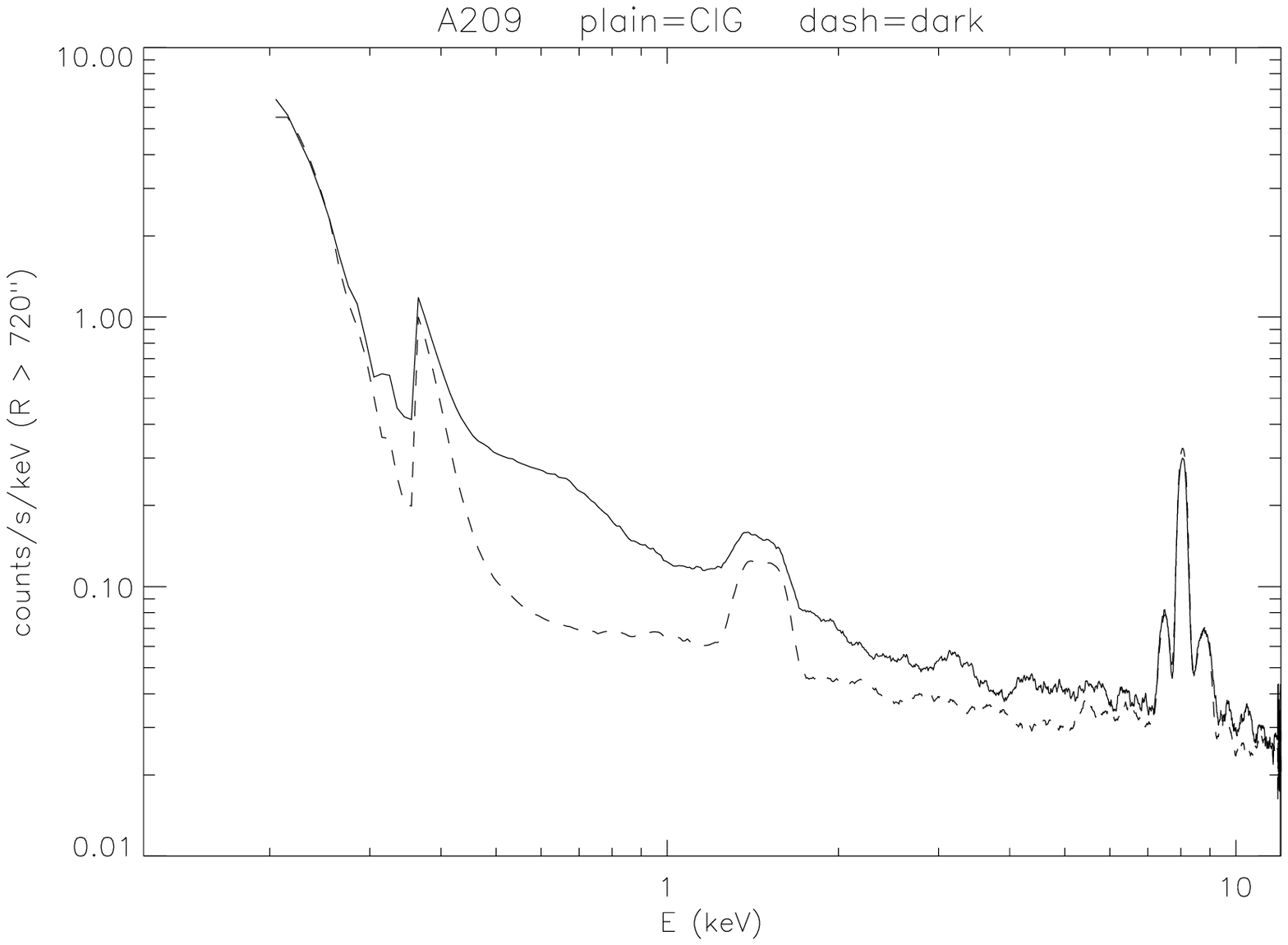,width=0.45\textwidth}
\caption{\label{fig:closed}LEFT: comparison of the spectra from dark fields as seen by \mos1, \mos2 and \pn, in their full FOV. RIGHT: comparison of the spectra from an outer free region of the \pn\ \abel\ observation (REV 202) and the corresponding region in the renormalized dark data; the difference is the local EXRB.}
\end{center}
\end{figure}

	A first study of dark fields was conducted in January 2002. At that time, using the online log browser from the \xmm\ internet site, a list of data sets could be established: c.a. $190~ks$ for full frame \mos1, $170~ks$ for full frame \mos2, $114~ks$ for full frame \pn, $30~ks$ for extended full frame \pn. But not all of them were available from the public FTP (part of proprietary data, or not yet processed by the official pipeline which treats science observation in priority), and from those available a few appeared contaminated by those periods of unusually high overall background level (\cha{sec:flare}). After processing, only remained respectively\footnote{corresponding data files have been made public, and will be regularly maintained and updated, on \verb!ftp://www-station.ias.u-psud.fr/pub/epic/Closed!}: $110~ks$, $101~ks$, $15~ks$ and $27~ks$. Dark data for other modes were even rarer and at any rate beyond the scope of this first analysis, for ClG observations seldomly make use of them.

	As shown on \fig{fig:closed}, dark fields contain all background components except soft protons and the EXRB. As for blank fields, they are dependent on the instrumental mode (\pn\ full frame and extended frame count rates differ by a factor of about about $1.4$ in the full FOV and in the $100:7100~keV$ range; the factor goes down to $1$ in the $1100:7100~keV$ range), but are easier to merge together since they require no particular cleaning (neither flares nor point sources). During the last months, new observations in closed filter have been performed and will soon be added to this analysis to improve statistics.

  \subsection{Software upgrades} 

	A last caveat to be aware of is the spectral discrepancy that may appear when using blank or dark fields, that have been processed with a different \sas\ software version than for the main data. Indeed, algorithms taking CTI effects (\cha{sec:intro}) into account have been regularly improved, resulting in better event energy reconstruction and hence more accurate fluorescence line centering. Other parameters, like the effective exposure duration, may also gain in accuracy as the \sas\ chain tasks and CCF (Current Calibration Files) improve with time, so that on one hand users should be encouraged to process their data with the latest available softwares, and on the other hand, blank and dark merged data sets should be maintained for consistency.

\section{SPECTRO-IMAGING ANALYSIS}
\label{sec:spectro}

	Considering that now instrumental and particles background components may be filtered out or corrected for satisfactorily enough, the last step of analysis should yield to the EXRB correction. Again, different methods may be tried out, depending mainly on the size of the studied source relatively to the \epic\ FOV.

  \subsection{Field source removal} 
\label{sec:sources}

	As already mentioned in \cha{sec:skybkg}, particular care must be given to the detection and masking off of field sources that may contaminate the diffuse emission from the analysed extended source.

\begin{figure}[h!bt]
\begin{center}
\hspace{0cm}
\psfig{figure=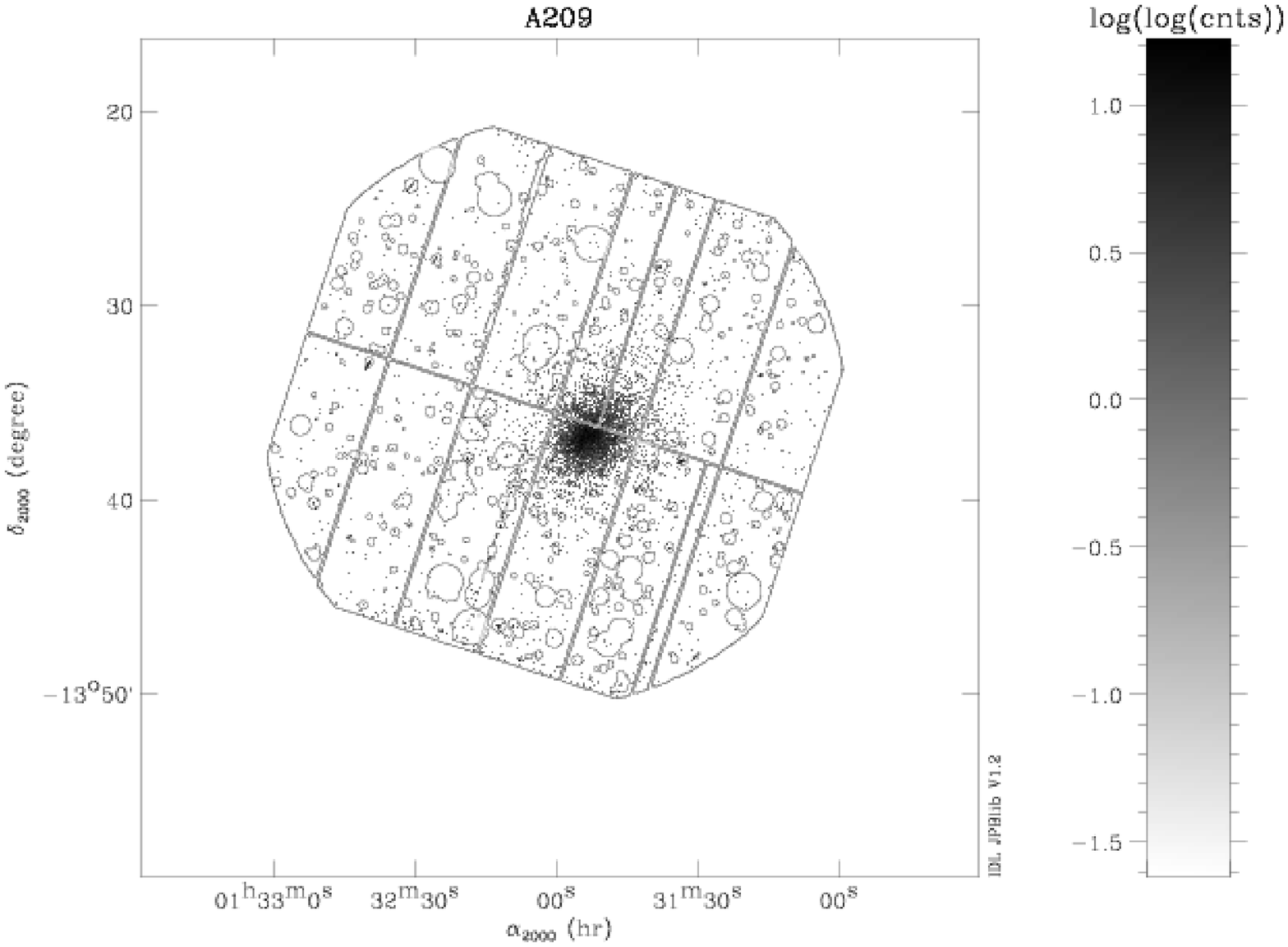,width=0.45\textwidth,angle=90}
\psfig{figure=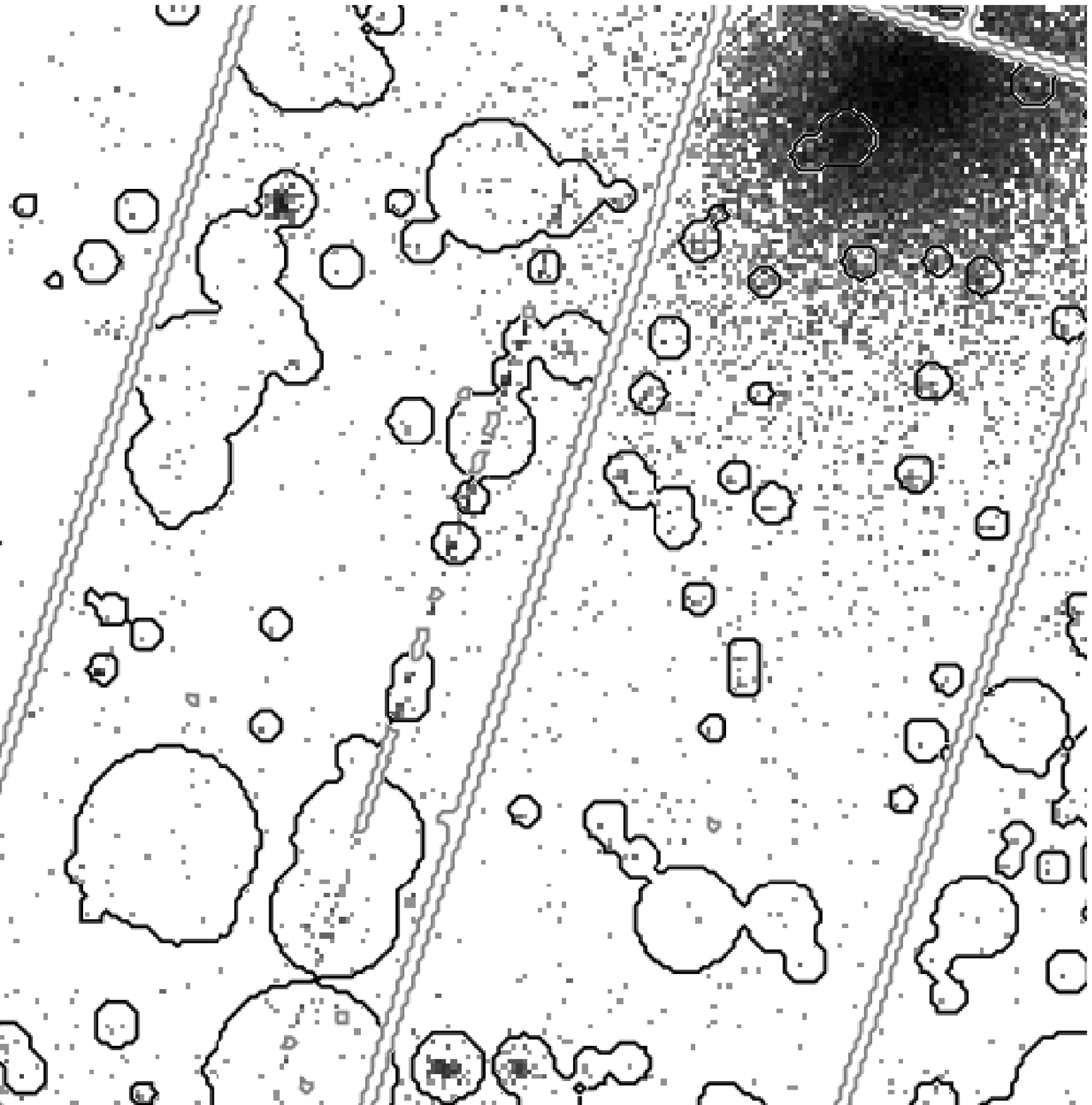,width=0.45\textwidth}
\caption{\label{fig:sources}LEFT: result of the $ewavelet$ \sas\ task (from scales $1$ to $32$, threshold $3 \sigma$, $2$ iterations) applied to the \pn\ image of \abel\ (REV 202); black contours are detected sources and bright spots, gray contours are the CCD mask and the badpixels as detected by the standard \sas\ $epchain$ pre-analysis task. RIGHT: zoom on a bright column, not completely masked by $epchain$.}
\end{center}
\end{figure}

	At the moment, two different detection tasks are provided within the \sas\ (\verb!emldetect! and \verb!ewavelet!) using different algorithms and achieving different performance\cite{bib:Brunner01}, but other methods are also under development\cite{bib:Valt01,bib:Bourdin01,bib:Rasmus01}. The \verb!ewavelet! tasks has nevertheless this particularity of being able to detect remaining spurious bright pixels and columns (\fig{fig:sources}), which may be a drawback when searching for serendipitous new astrophysical sources, but an advantage in the purpose of cleaning the emission from a known extended object. Then, the problem of deciding whenever any source detection really is a separated foreground (or background) source or part of the diffuse emission structure araise. In the case of ClG observations, the straightforward method consists of masking off any detected bright spot, but that point would deserve a serious investigation in the light of the more and more numerous reports about cooling flows and substructures origins in the intracluster medium ({\it cf.} proceedings from dedicated conferences).

  \subsection{The weights method} 
  \label{sec:weights}

	A last caveat about extended source data analysis concerns the mixing of spatial and spectral instrumental artifacts, so that it is difficult to extract correct images or spectra from X-rays event lists. The standard way to correct an image for mirror vignetting effect is to divide it by a ``flat field map'' (or ``exposure map'' when it is multiplied by the integration time), but spectral information about energy dependent efficiency is lost. The same for a spectrum which is in general compared to a model folded through the instrument response (RMF and ARF), only valid for a given off-axis angle (in general $\theta=0$), while the spatial off-axis dependent vignetting effect may extend over a substantial fraction of the detector. 

	Spectroscopy is one key investigation tool for intracluster gas temperature structure and physics, that is why it has been proposed\cite{bib:Arnaud01b}, in the frame of very extended sources studies, to correct spectra for spatial vignetting effects using a ``weights'' method inherited from {\it ROSAT} experience. The basic principle is to accumulate, for each listed event of energy $E$ and off-axis position $\theta$, $N$ effective events instead of only $1$, where $N = 1 / V(E,\theta)$ (\eqn{eqn:mirror}). This method has already been extensively described\cite{bib:Arnaud01b,bib:Maj02}, as well as the possible generalization ($N = 1 / ARF(E,\theta)$, \eqn{eqn:ARF}) which could be used to correct in one pass images for spectral features as well as vignetting\cite{bib:Marty01,bib:Marty02}, and has been eventually implemented in the last \sas\ software version\footnote{The user should be warned, as very recently noticed at the last EPIC-CAL meeting, that this new {\verb!evigweight!} task produces wrong output when used in conjunction with a version of the {\verb!XMM\_MISCDATA!} CCF older than {\verb!0014!}} (\verb!v5.3!).

	This method is in addition fully compatible with the background removal methods using auxiliary data (\cha{sec:skybkg},\cha{sec:detbkg}) as long as the same detector areas are considered in each data set (by masking off the same detected point sources regions in the auxiliary set for instance), since non-vignetted background events will be abusively weighted in the same way within both sets and will still cancel out by subtraction. On the other hand, this method increases the noise on the results, because of the uncertainty on the individual weights coming from the uncertainty on the individual positions (smeared by the PSF) and individual energies (smeared by the RMF).

  \subsection{Brightness Profiles} 
  \label{sec:prof}

	Brightness profiles are usually extracted to study the intracluster mass profile, within relaxed ClG (where spherical symmetry may be assumed). As the \verb!radial! \sas\ task has now disappeared from the new version release, due to maintainance difficulties, the general purpose \verb!evselect! event selection routine should be prefered, asking for a radial distance histogram; a radial distance column may be easily computed on the basis of the $X$ and $Y$ columns which are relative to the ``RA-DEC'' WCS (Right Ascension - Declination World Coordinate System) using the \verb!tabcalc! for example. More accurate routines, for computation of true angular distance along sky great circles, have been developped using {\it IDL} softwares\footnote{they will be made public as soon as possible}, and probably are also available in packages like {\it EXSAS}\footnote{\verb!http://www.rosat.mpe-garching.mpg.de/~web/exsas.html!}.

	In the light of the previous sections, it should now be possible to subtract adequately a corresponding background profile to that coming from the main data. But problems raised earlier shall now become apparent.

	Using only blank fields for instance, in combination to the weights method for a very extended source like the Abell~2163 ClG, may lead to some residuals, especially at soft energies\cite{bib:Pratt01}. Furthermore, in case the main data have been acquired using a different filter than for blank fields, the generalized weights method should be applied (correcting not only for spatial vignetting, but also for spectral transmission, at least that of the filters), but then the weights will not be the same for both data sets causing instrumental background subtraction bias. One can call twice upon the dark fields data set in order to clean the main and blank fields separately, before extracting radial profiles and subtract them, but this implies many more operations and memory usage, and increases the noise.

\begin{figure}[h!bt]
\begin{center}
\hspace{0cm}
\psfig{figure=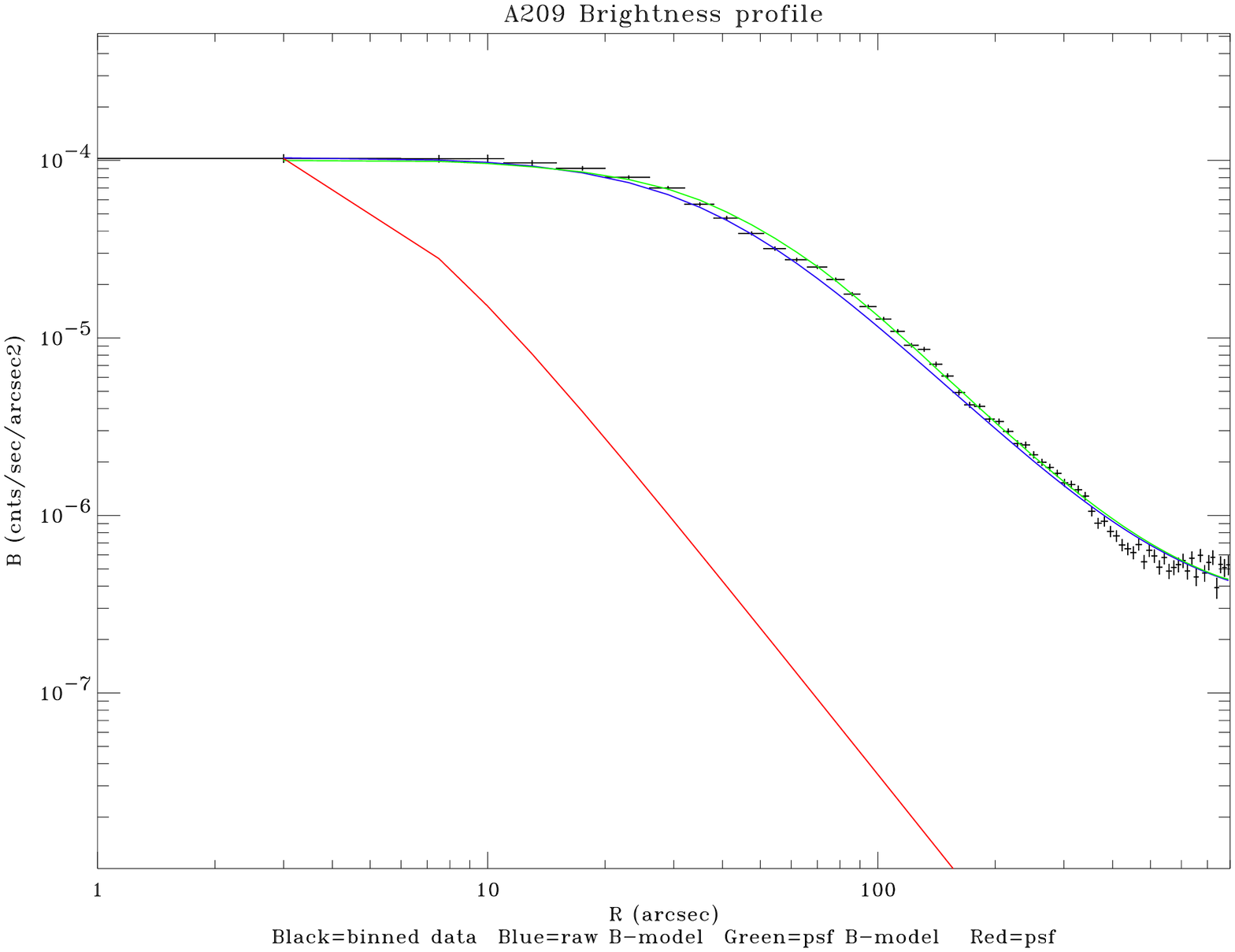,width=0.45\textwidth,angle=90}
\psfig{figure=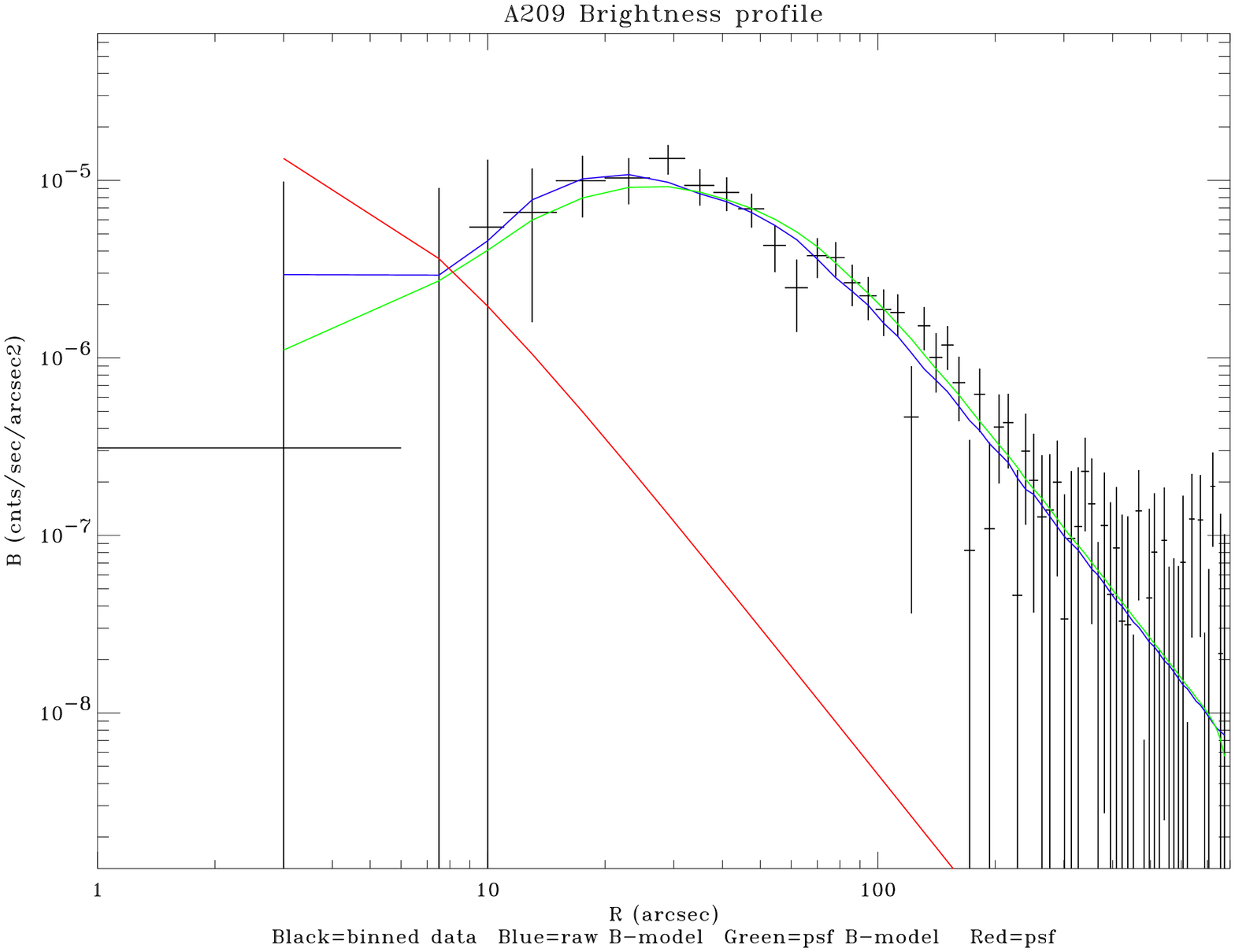,width=0.45\textwidth,angle=90}
\caption{\label{fig:radprof}LEFT: cumulated \xmm\ \epic\ (\pn+\mos1+\mos2) radial profile of \abel\ (REV 202), fitted by PSF-convolved $\beta$-model ($R_c=38~arcsec$, $\beta=0.53$). RIGHT: \xmm\ \epic\ cumulated differential profile with the same PSF-convolved $\beta$-model that has been differentiated and overplotted.}
\end{center}
\end{figure}

	Another way is to use only dark fields to remove instrumental noise from the main data, leaving only the EXRB in the resulting radial profile (\fig{fig:radprof}). It can be then included in the model to be fitted as an additional constant parameter\cite{bib:Marty02}. If this is not desirable for any reason, the EXRB may also be estimated from an outer free region in the main data, before being subtracted from it (or added to model). Current EXRB modelization attempts\cite{bib:Gilli01} may also provide a useful value if the main source is too extended to leave empty regions on the detector. Finally, a last method consists of computing a differential radial profile (where the $i$th bin of the differential profile $dP$ may be deduced from the original profile $P$ bins by $dP_i = P_i - P_{i+1}$), which may be simply fitted by a differential model, where any constant term (including the EXRB) should have been canceled out in the differentiation process (\fig{fig:radprof}).

	Note that any model to be fitted to the profile should be first convolved by the PSF. This step is not trivial since the PSF differs slightly from one \xmm\ telescope to another, and also is a function of energy and off-axis angle. A temporary method consists of computing a kind of average PSF over the considered detector region and energy band, according to the most recent calibration\cite{bib:Ghiz02}. A more sophisticated method is discussed in \cha{sec:maps}. The only calibration parameter that has not been taken into account at this point is the RMF (the pure redistribution effect; the spectral efficiency may be included in the generalized weights if necessary; see \cha{sec:detec},\cha{sec:weights}), but it should not be an issue as long as the radial models do not truly depends on energy.

  \subsection{Wide Spectra} 
  \label{sec:spec}

	Spectra are directly extracted using the \verb!evselect! task. The same kind of cleaning algorithms as in \cha{sec:prof} may be used, and the same difficulties may be encountered. Most spectral fitting packages, like {\it XSPEC}\footnote{\verb!http://heasarc.gsfc.nasa.gov/lheasoft!}, are designed to handle different data files which account for various background components, as well as instrumental response (ARF \& RMF, \cha{sec:detec}). The main problem here is to take the PSF into account whenever extracting spectra from narrow regions, like concentric annuli for the purpose of temperature profile reconstruction\cite{bib:Mark02}. But as long as the region is more extended than the PSF \cha{sec:detec}, this should not be an issue.

  \subsection{Hardness Profiles and Maps} 
  \label{sec:maps}

	But since \xmm\ opened the spatially resolved spectroscopy window (\cha{sec:intro}), all the previous methods may appear somewhat restricted in the frame of studying the intracluster medium structure. That is the reason that motivated the following new methodology. Its driving principle resides in that PSF corrections may be only carried out within images, and that if the energy band of those images was large enough, the RMF should not be an issue.

\begin{figure}[h!bt]
\begin{center}
\hspace{0cm}
\psfig{figure=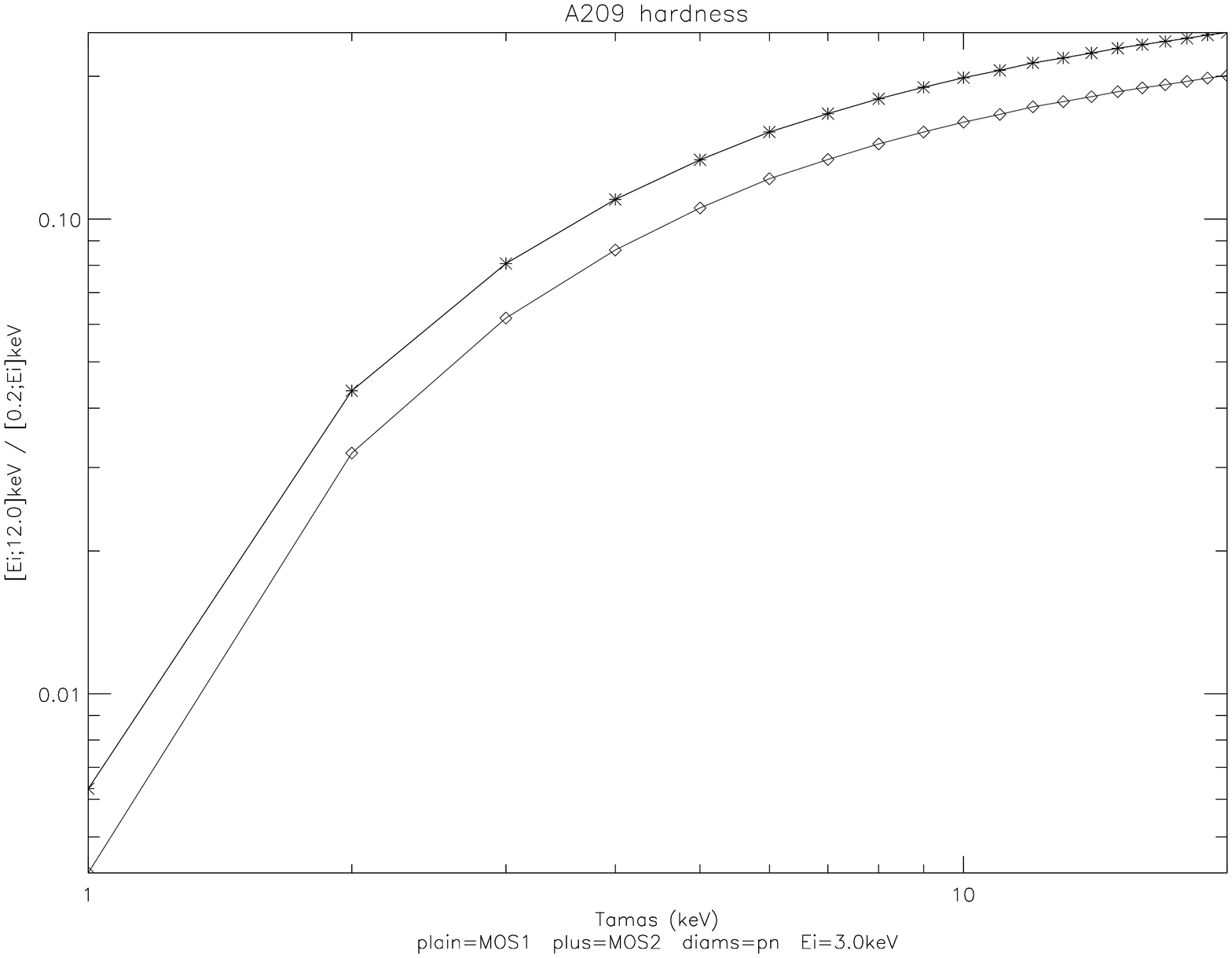,width=0.45\textwidth,angle=90}
\psfig{figure=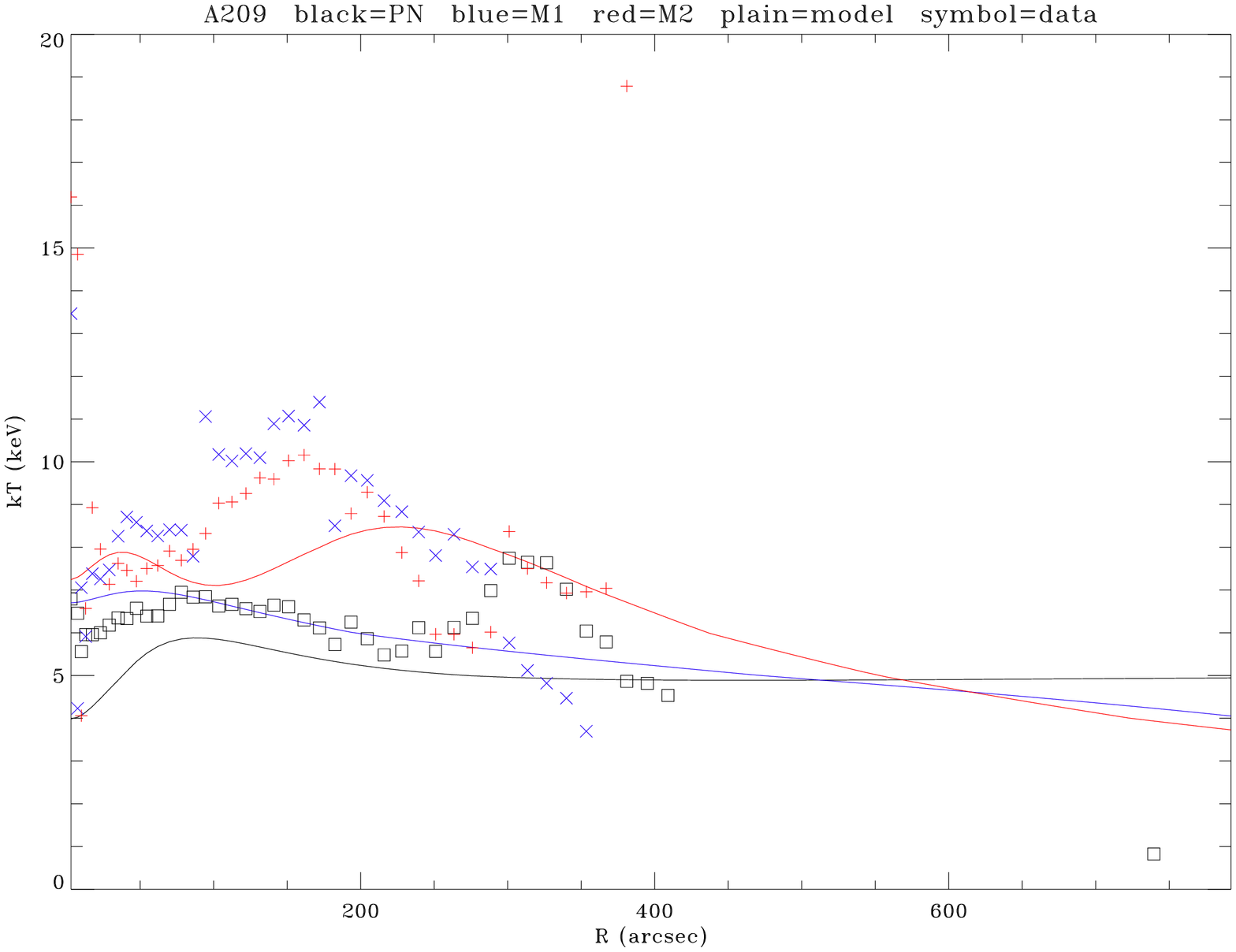,width=0.45\textwidth,angle=90}
\caption{\label{fig:hrprof}LEFT: example of hardness-to-temperature conversion curve, for the \pn\ and \mos\ cameras, with the hardness definition $[0.2:3.0~keV] \over[3.0:12.0~keV]$. RIGHT: \pn, \mos1 and \mos2 tentative temperature profiles of \abel\ (REV 202), with the raw data profiles in symbols and the PSF-deconvolved profiles in plain lines.}
\end{center}
\end{figure}

	And, indeed, it is possible to get a temperature measurement through the use of images, extracted in two (or more) different energy bands and recombined to form a hardness image, just as for ``false color images''. Classical combinations use two images, one ``hard'' (in a high energy band) image $H$ and one ``soft'' (in a low energy band) image $S$, and are of the form $H / S$ (which is simple) or $(H-S) / (H+S)$ (which limits divergence when fluxes drop to zero). A hardness-to-temperature conversion function may be build using a spectral model fitting software (\fig{fig:hrprof}), by calculating the hardness ratio from model spectra at different temperatures.

	A first approach consists then in extracting two radial brightness profiles in two energy bands, and building a hardness profile using previous relations, which can then be translated into temperature units. The hardness profile should rather be build from models fitted to the brightness profiles (\cha{sec:prof}) in order to deconvolve from the PSF effect. A tentative example is shown in \fig{fig:hrprof}, but without error bars yet due to the youth of the method. More accuracy has even been obtained by fitting models to brightness profiles within narrow energy bands ($\Delta E = 500~eV$), and recombining them afterwards into larger bands ($0.2:3.0~keV$ and $3.0:5.0~keV$).

\begin{figure}[h!bt]
\begin{center}
\hspace{0cm}
\psfig{figure=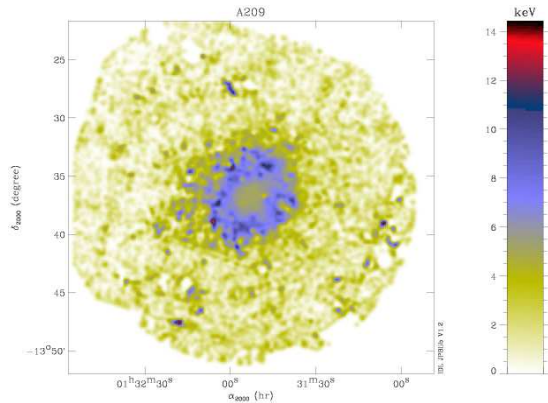,width=0.45\textwidth,angle=90}
\caption{\label{fig:hrmaps}Cumulated \pn, \mos1 and \mos2 tentative temperature map of \abel\ (REV 202)}
\end{center}
\end{figure}

	The next step is to apply the same algorithm to images, which can also be splitted into narrow energy bands, and even annuli (at the cost of heavier memory and processor duty). Then a smoothing algorithm using a kernel equal to the PSF core may provide a way to correct the images from spatial smearing effects (\fig{fig:hrmaps}). Better, a two-dimensional brightness model, convolved by the PSF, could be fitted to the maps, allowing a hardness determination based on those models; the drawback still being that this forbids any substructure analysis that has not been included in the models {\it ab initio}. Eventually, a temperature image may be reconstructed using the above method.

	However, all this is obviously very idealistic, since the correct methodology would be in fact to reverse-engineer the events through the PSF, but also the RMF since it is now question of splitting data into narrow energy bands. In addition, such energy and off-axis splits lead to regions with very poor statistics, and global methods should be prefered. Methods are indeed currently investigated, on the basis of data inversion algorithms\cite{bib:Tegmark97} generally used for millimetric data coming from bolometers to study anisotropies in the diffuse Cosmological Microwave Background, or on the basis of multiscale wavelet image restoration algorithms\cite{bib:Bourdin01}. While the former is entirely based on instrumental response and seems very promising, the latter makes use of very little {\it ad hoc} hypothesis but it is not yet clear how it can properly handle PSF, and above all RMF, aspects.

\section{CONCLUSIONS} 

	This review lists the various analysis methods required by extended sources data from \xmm\ \epic\ instruments. It emphasizes the difficulties of automated batch processing (in the frame of surveys or catalogs of ClG), and the need for improving and maintaining tools (for weights correction, profiles and map extraction\ldots) and auxiliary data files, like the blank fields and dark fileds which could include more statistics and be available for all instrumental settings (mode and filter). At least three teams\cite{bib:Lumb01,bib:Marty01,bib:Ghiz00} actively work on that latter topic, while the core of the \sas\ softwares is managed by the \xmm\ SOC\footnote{\verb!http://xmm.vilspa.esa.es!}.

	In addition, more studies are still needed concerning the modelization of the EXRB, which is also a hot topic\cite{bib:Ghiz00,bib:Gilli01,bib:Marty02}, as well as in the field of foreground or background source detection algorithm (\cha{sec:sources}).

	The main difference with point source objects analysis is eventually held in the background determination and subtraction problem, in all its aspects, since extended sources suffer from the dilution of their flux at levels sometimes not much higher than the background itself. The farther the object, the fainter its flux and the more critical the background problem; on the other hand, the closest the object, the more it fills the FOV leaving little free detector area to estimate the background. And using even inaccurate auxiliary data may still be an issue in the determination of the true extension and mass of extended objects, which border regions are fainter than their cores.

	The ultimate goal of this work is to make batch analysis possible for ClG data, eventually pipelining the softwares to online internet databases, like BAX\cite{bib:Sadat02} or the \xmm\ science archive.




\acknowledgments     

This work is based on observations obtained with \xmm, an {\bf ESA} science mission with instruments and contributions directly funded by {\bf ESA} Member States and the {\bf NASA}.

The authors wish to thank Alain Blanchard, David Lumb, Bruno Altieri, Richard Saxton, Christian Erd, as well as the whole ``EPIC-CAL'' team.



\end{document}